\documentclass[conference]{IEEEtran}
\IEEEoverridecommandlockouts
\usepackage{cite}
\usepackage{amsmath,amssymb,amsfonts}
\usepackage{mathtools}
\usepackage{algorithmic}
\usepackage{graphicx}
\usepackage{textcomp}
\usepackage{xcolor}
\usepackage{caption}
\usepackage{subcaption}
\def\BibTeX{{\rm B\kern-.05em{\sc i\kern-.025em b}\kern-.08em
    T\kern-.1667em\lower.7ex\hbox{E}\kern-.125emX}}
\begin{document}

\title{Automated Movement Detection with Dirichlet Process Mixture Models and Electromyography\\
}

\author{Navin Cooray$^{1}$, Zhenglin Li$^{2}$, Jinzhuo Wang$^{3}$, Christine Lo$^{3}$, Mahnaz Arvaneh$^{2}$, Mkael Symmonds$^{4}$, \\Michele Hu$^{3}$, Maarten De Vos$^{5}$, and Lyudmila S Mihaylova$^{2}$% <-this % stops a space
\thanks{$^{1}$Institute of Biomedical Engineering, University of Oxford, Oxford, United Kingdom
        {\tt\small navin.cooray@ndcn.ox.ac.uk}}%
\thanks{$^{2}$Department of ACSE, University of Sheffield, Sheffield, United Kingdom 
        }%
\thanks{$^{3}$Nuffield Department of Clinical Neurosciences, University of Oxford, Oxford, United Kingdom
        }% 
\thanks{$^{4}$Department of Clinical Neurophysiology, University of Oxford, Oxford, United Kingdom
        }%  
\thanks{$^{5}$Department of Electrical Engineering, Catholic University of Leuven, Leuven, Netherlands
        }%          
}

\maketitle

\begin{abstract}
Numerous sleep disorders are characterised by movement during sleep, these include rapid-eye movement sleep behaviour disorder (RBD) and periodic limb movement disorder. The process of diagnosing movement related sleep disorders requires laborious and time-consuming visual analysis of sleep recordings. This process involves sleep clinicians visually inspecting electromyogram (EMG) signals to identify abnormal movements. The distribution of characteristics that represent movement can be diverse and varied, ranging from brief moments of tensing to violent outbursts. This study proposes a framework for automated limb-movement detection by fusing data from two EMG sensors (from the left and right limb) through a Dirichlet process mixture model. Several features are extracted from 10 second mini-epochs, where each mini-epoch has been classified as 'leg-movement' or 'no leg-movement' based on annotations of movement from sleep clinicians. The distributions of the features from each category can be estimated accurately using Gaussian mixture models with the Dirichlet process as a prior. The available dataset includes 36 participants that have all been diagnosed with RBD. The performance of this framework was evaluated by a 10-fold cross validation scheme (participant independent). The study was compared to a random forest model and outperformed it with a mean accuracy, sensitivity, and specificity of 94\%, 48\%, and 95\%, respectively. These results demonstrate the ability of this framework to automate the detection of limb movement for the potential application of assisting clinical diagnosis and decision-making. 

% Research continues to uncover the relationship between neurology and sleep, one such example is the strong  capability of RBD to predict Parkinson's disease. The diagnosis of RBD can predict the onset of Parkinson's by years, potentially decades.

\end{abstract}


\section{Introduction}
%Intro
Ongoing research into sleep continues to highlight its significance to mental and physical well being \cite{Grandner2017}. Studies of numerous sleep disorders appear to preempt the onset of numerous neurological disorders. This includes rapid-eye movement (REM) sleep behaviour disorder (RBD), where mounting evidence suggests that this parasomnia predicts Parkinson's disease (PD) by years, potentially decades \cite{Schenck2013b,Postuma2019}. This predictive ability provides an opportunity to explore preventative medicine and better understand how neurodegenerative disorders develop over time. PD is the second most prevalent neurodegenerative disease worldwide, affecting more than four million people \cite{Bovolenta2017}. Beyond the major impact to quality of life and increased mortality, the chronic nature and growing disability of PD incurs major healthcare expenses that will only continue to escalate in countries with an ageing population \cite{Ballesteros2016,Winter2010}. More work is required to understand the development of this disorder so that preventative measures can be devised. RBD represents one potentially promising early predictor for a large part of PD sufferers, possibly providing a clear avenue to target remedies before the onset of PD.

Characteristic muscle activity associated with RBD includes complex and simple limb movements. For sleep studies, limb movement activity is captured using EMG sensors, which are within the electrostatic categorisation of sensing technology \cite{Fu2020}. Clinicians are taught to visually identify EMG activity without a clear and precise definition. Visually identifying muscle activity to describe limb movement is also critical in diagnosing restless leg syndrome (RLS) and periodic limb movement disorder (PLMD). RLS has been found to be one of the most common sleep disorders in the United States of America \cite{Winkelman2006,Allen2005}. One study suggests RLS and PLMD are associated with cardiovascular disease and hypertension \cite{Walters2009}, while another has found a link between secondary RLS (occurs secondary to other medical conditions) and cardiovascular disease \cite{VanDenEeden2015}.

The AASM has defined RLS as an urge to move the legs, which must begin or worsen at rest, be partially or totally relieved when in movement, and occurs predominantly at night \cite{Iber2007}. These movements must not by accounted for by another conditions such as leg cramps, arthritis, or positional discomfort. PLMD is far less common and is characterised by periodic episodes of repetitive limb movement during sleep and is distinct from RBD or RLS. 

With the prevalence of sleep disorders continually increasing and the growing demand to better understand sleep and its implications on physiology (for example in RBD and RLS), the burden placed on sleep clinics is great and their efforts often hampered by manually-laborious diagnostic procedures. As a result researchers are keen to explore the viability of automated diagnostic support-tools to increase efficiency, accuracy, and productivity. Furthermore, the utility of automated sleep analysis, provides the opportunity to better understand sleep and its association with neurodegenerative and cardiovascular diseases. 

The rest of this paper is organised as follows. Section II presents an overview of related work, Section III details the problem formulation for the automated movement detection with Dirichlet process models and how fusion of EMG data from the left and right leg movement is performed.

\section{Related Work}
Numerous studies aim to provide automated techniques to identify various sleep disorders, sleep stages, or even specific sleep characteristics. A select few algorithms look at automating the detection of abnormal movement during sleep using EMG signals from the chin. The AASM stipulates at least a single EMG sensor to be placed on the chin in order to clinically analyse sleep and specifically identify abnormal muscle movement \cite{Iber2007}. Diagnosing bruxism requires the evidence of teeth grinding during sleep, as such a few studies exist detailing a portable device to detect bruxism episodes \cite{Castroflorio2013,Castroflorio2014}. These two studies focused on using a simple EMG amplitude threshold in combination with heart rate elevation (measured from an ECG sensor) to identify bruxism episodes. This study demonstrated the predictive ability of an algorithm to aid in identifying bruxism, however the degree of variation and complexity of sleep disorder movements would mean that a simple threshold would not suffice for applications in PLMD and RBD. As a result the concept of automated movement detection algorithm lends itself towards a non-parametric model that can incorporate numerous sensors and compensate for movement which can vary greatly in magnitude and severity. A handful of other studies demonstrate this through limb movement detection in participants with RBD and PLMD \cite{Cesari2018a,Cesari2019a,Cesari2019}.

In one study, Cesari \textit{et al.} (2018), demonstrated the utility of a non-parametric probabilistic model to distinguish leg-movement from resting EMG mini-epochs \cite{Cesari2018a}. From a dataset containing $27$ healthy controls and $36$ participants diagnosed with PLMD, this study was able to utilise this semi-supervised approach to detect PLMD participants with $82$\%. As an extension of this study, Cesari \textit{et al.} (2019), applied this technique on a mixed cohort of $27$ healthy controls, $36$ individuals diagnosed with PLMD, and $29$ participants diagnosed with RBD \cite{Cesari2019}. While these studies didn't explore the performance of limb-movement detection (as manual annotations of limb movement are rare), it did validate its utility in distinguishing RBD and PLMD participants from healthy individuals. In a follow-up study this technique was expanded to a German sleep study that was able to assess the performance of limb-movement detection through the PLMS-index \cite{Cesari2019a} using three EMG sensors (from the chin, left tibia, and right tibia). This German dataset contained $240$ participants that were healthy controls or diagnosed with combinations of PD, PLMD, and RBD \cite{Cesari2019a}. Each participant was given a PLMS-index score which details the average number of limb movements per hour of sleep. Using the aforementioned techniques \cite{Cesari2018a,Cesari2019}, this study demonstrated an automated classification of participants with PLMD and RBD with an accuracy of $88.75$\% and $84.17$\%, respectively \cite{Cesari2019a}. Once more this study was able to assess the performance of limb movement detection by achieving an automated PLMS-index score that correlated to the manual score by $84.99$\% and only had a slight bias towards over-predicting the PLMS-index \cite{Cesari2019a}. However, these studies are limited in that they provide a proxy to individual event detection of limb movement (the PLMS-index) without exploring the limb-movements as seen or annotated by sleep clinicians. 
Nonetheless these studies have demonstrated the utility and potential of limb movement detection in the automated identification of specific sleep disorders. 

\section{Problem Formulation}
In a previous study, Li \textit{et al.} 2020 demonstrated the utility of a Dirichlet Process (DP) mixture model to automate the detection of sleep apnea segments and motivated movement detection in this study \cite{Li2020}. The advantage of this framework is in the data-driven approach to learn number of clusters within the mixture models. The DP is defined as a distribution over distributions \cite{Gershman2012}. Namely, where each observation of $x_i$ is generated from a distribution with parameter(s) $\boldsymbol{\theta_i}$, which itself is generated from a prior distribution $G$:

\begin{align}
& \boldsymbol{\theta_i}\mid G \sim G &\text{ for each }i\label{eq1}\\
& x_i\mid\boldsymbol{\theta_i} \sim F(\boldsymbol{\theta_i}) &\text{ for each }i,\label{eq2}
\end{align}
where $F(\boldsymbol{\theta_i})$ is the distribution of $x_i$ given parameter(s) $\boldsymbol{\theta_i}$ (note that differing $\boldsymbol{\theta_i}$s are not necessarily distinct values).  

Consider a measurable space and any finite partitions $\{T_1,....,T_K\}$ of it. If $G \sim DP(\alpha,G_0)$, then:
\begin{equation}
(G(T_1),...,G(T_K)) \sim \text{Dir}(\alpha G_0(T_1),...,\alpha G_0(T_K)). \label{eq3}    
\end{equation}
where $G_0$ is defined as the base distribution with a concentration parameter $\alpha$.

The DP can be constructed by considering a unit length stick that is divided into an infinite number of segments represented by $\pi_k$, in the following manner:
\begin{align}
& \beta_k \sim \text{Beta}(1,\alpha) \label{eq4}\\
& \pi_k = \beta_k\prod_{j=1}^{k-1}(1-\beta_j) = \beta_k\left(1-\sum_{l=1}^{k-1}\pi_l\right). \label{eq5}
\end{align}
where $\boldsymbol{\pi}=\{\pi_k\}_{k=1}^{\infty}$ is a sequence of mixture weights and $k$ denotes the index of the component. Finally a DP is constructed in the following way:
\begin{align}
& \boldsymbol{\theta_k^*} \sim G_0 \label{eq6}\\
& G = \sum_{k=1}^{\infty}\pi_k\delta_{\theta_k^*} \label{eq7}\\
& G \sim \text{DP}(\alpha,G_0), \label{eq8}
\end{align}
where $\{\theta_k^*\}_{k=1}^\infty$ are independent
and identically distributed (i.i.d.) random variables drawn from the base distribution $G_0$ along with draws for weights ($\pi_k$) as expressed in \eqref{eq5}.

Consider features extracted from the $i$-th segment as $x_i$, its distribution can be expressed as follows:
\begin{equation}
p(x_i) = \sum_{k=1}^{K}\pi \mathcal{N}(x_i;\boldsymbol{\theta_k^*}), \label{eq9}    
\end{equation}
where $\mathcal{N}(.)$ denotes the Gaussian distribution and the parameters of the $k$-th component are denoted by $\boldsymbol{\theta_k^*} \overset{\Delta}{=} \{\boldsymbol{\mu_k^*},\boldsymbol{\Sigma_k^*}\}$. The mean vector and variance matrix of the $k$-th Gaussian component are represented by $\boldsymbol{\mu_k^*}$ and $\boldsymbol{\Sigma_k^*}$, respectively. 

Mixture model theory assumes that each $x_i$ is generated by first choosing a cluster, indexed by an assignment variable $z_i$ according to a categorical distribution of $\boldsymbol{\pi} = [\pi_1,...,\pi_K]$ \cite{Gershman2012,Li2018}. The $x_i$ observations are then generated from the chosen component with the parameter $\boldsymbol{\theta_i}=\boldsymbol{\theta}_{z_i}^*$. Because the number of components, $K$, and the distribution weights, $\boldsymbol{\pi}$, are unknown and are solved using the available observsations. The framework of the DP allows to solve this problem and when combined with the stick-breaking process (detailed before) the generative model can be described as follows:
\begin{align}
& z_i \sim \boldsymbol{\pi} \label{eq10}\\
& x_i \sim \mathcal{N}(\boldsymbol{\theta}_{z_i}^*), \label{eq11}
\end{align}
where $\{\boldsymbol{\theta}_k^*\}_{k=1}^{\infty}$ are distinct values of the parameters $\boldsymbol{\theta}_k^*$s, sampled independently from the base distribution $G_0(\boldsymbol{\theta}^*\mid\lambda)$ (detailed in \eqref{eq6}, where $\lambda$ is the hyperparameter of $G_0$) and the distribution of $\boldsymbol{\pi}$ is given in \eqref{eq5}.

Suppose the parameters $\boldsymbol{\theta_{k}^{*}}$s and $\beta_k$s are denoted as $\boldsymbol{\Theta} = \{\boldsymbol{\theta_{k}^*}\}_{k=1}^\infty$ and $\boldsymbol{\beta}= \{\beta\}_{k=1}^\infty$, respectively. The random variables $\boldsymbol{\beta}$ are drawn independently from a Beta distribution as defined in \eqref{eq4}. Let $\boldsymbol{z}=\{z_i\}_{i=1}^N$ be the cluster assignments of $N$ training features $\boldsymbol{X}=\{x_i\}_{i=1}^N$ and $\boldsymbol{W}=\{\boldsymbol{\beta},\boldsymbol{\Theta},\boldsymbol{z}\}$ be the collection of all latent parameters. Often in clustering problems the predictive density is calculated, and given the features $\boldsymbol{X}$ for training and a new sample $x'$ for testing, the probability of $x'$ being generated from the trained model can be expressed using the product-rule:

\begin{align}
    &p(x'\mid\boldsymbol{X})\notag\\&=\int p(x'\mid z',\boldsymbol{W},\boldsymbol{X})p(z'\mid \boldsymbol{W},\boldsymbol{X})p(\boldsymbol{W}\mid \boldsymbol{X})dz'd\boldsymbol{W}  \label{eq12}\\
    &=\int p(x'\mid z',\boldsymbol{\beta},\boldsymbol{\Theta},\boldsymbol{z},\boldsymbol{X})p(z'\mid \boldsymbol{\beta},\boldsymbol{\Theta},\boldsymbol{z},\boldsymbol{X})p(\boldsymbol{W}\mid \boldsymbol{X})dz'd\boldsymbol{W} \label{eq13}\\
    &=\int p(x'\mid z',\boldsymbol{\Theta})p(z'\mid \boldsymbol{\beta})p(\boldsymbol{W}\mid \boldsymbol{X})dz'd\boldsymbol{W}\label{eq14}\\
    &=\int p(x'\mid \boldsymbol{\theta_{z'}^{*}})p(z'\mid \boldsymbol{\beta})p(\boldsymbol{W}\mid \boldsymbol{X})dz'd\boldsymbol{W} \label{eq15}   
\end{align}
where $z'$ is the cluster assignment of the testing data $x'$. 
From \eqref{eq15} we can observe the first term, $p(x'\mid\boldsymbol{\theta_{z'}^{*}})$, can be calculated from \eqref{eq9} and \eqref{eq11}, while the second term, $p(z'\mid\boldsymbol{\beta})$ can be solved by \eqref{eq5} and \eqref{eq10}. However, the last term, $p(\boldsymbol{W}\mid \boldsymbol{X})$, is intractable but can be approximated using a variational distribution. A variational distribution is designed as a family of factorised distributions as described by meanfield variational inference \cite{Blei2017}:
\begin{align}
& q(\boldsymbol{W};\boldsymbol{\phi}) = \prod_{k=1}^{K}\big[q(\beta_k;\phi_k^\beta)q(\boldsymbol{\theta_k^*};\phi_k^{\theta^*})\big]\prod_{i=1}^{N}q(z_i)\label{eq16}
\end{align}
where $q(z_i)$s are categorical distributions, $\phi_k^\beta$ and $\phi_k^{\theta^*}$ are parameters of distributions of $q(\beta_k)$ and $q(\boldsymbol{\theta_k^*})$, with $\phi_k=\{\phi_k^{\beta},\phi_k^{\theta^*}\}$. Through variational inference these parameters are updated iteratively to find a minima, details of the derivation are detailed in \cite{Kurihara2007}. As a result \eqref{eq15} can be rewritten as:
\begin{flalign}
p(x'\mid\boldsymbol{X}) = &\int p(x'\mid \boldsymbol{\theta_{z'}^{*}})p(z'\mid \boldsymbol{\beta})q(\boldsymbol{W;\phi})dz'd\boldsymbol{W}\label{eq17}\\
= &\int p(x'\mid \boldsymbol{\theta_{z'}^{*}})p(z'\mid \boldsymbol{\beta})\prod_{k=1}^{K}\big[q(\beta_k;\phi_k^\beta)q(\boldsymbol{\theta_k^*};\phi_k^{\theta^*})\big]\notag\\&\prod_{i=1}^{N}q(z_i)\:dz'\:d\beta\:d\boldsymbol{\theta^*}\:d\boldsymbol{z}\label{eq18}
\end{flalign}
which can be calculated analytically. In this study, the DP Gaussian mixture model (DPGMM) was applied in the context of leg-movement detection in order to aid clinicians identify abnormal segments of sleep. 

Sleep medicine in its current form demands clinicians laboriously analyse polysomnography (PSG) recordings in order to make diagnostic decisions. These logistical bottle-necks often hinder epidemiological studies to better understand the link between sleep disorders and physiology, where RBD is just a single example. This study aims to utilise sleep recordings from RBD participants that contain annotated notes of limb-movement to assess a supervised probabilistic model of limb movement detection. 

\section{Polysomnography Data} \label{Data}

The John Radcliffe (JR) hospital retains PSG recordings as part of National Health Service (NHS) routine care for individuals suspected of having RBD. This study applied through the Clinical Trials and Research Governance (CTRG) to access anonymised case records for patients who were suspected of having RBD and later confirmed through these recordings. In addition to complete PSG data, these records included: age, sex, diagnosis (recorded by clinical staff) and treatment received at time of recording. PSG recordings were anonymised by those who had authority to access the data. In total $36$ participants were included in the PSG recordings and are summarised in Table \ref{table:jr_data}. This dataset provided two nights of full PSG recordings for each participant. Please note the male bias in the dataset, which is representative of the male predominance of RBD \cite{Schenck2002}. This study complied with the requirements of the Department of Health Research Governance Framework for Health and Social Care 2005 and was approved by the Oxford University hospitals National Health Service (NHS) Trust (HH/RA/PID 11957).

\begin{table}[htbp]
\caption{Dataset used for this study provided from the John Radcliffe hospital.}
\begin{center}
\begin{tabular}{|c|c|c|c|c|}
\hline
\textbf{Cohort} & \textbf{\#} & \textbf{Female} & \textbf{Male} & \textbf{Age (years)}\\
\hline
RBD Participants & 36          & 2               & 34             & $64.3\pm7.96$           \\
\hline
\end{tabular}
\label{table:jr_data}
\end{center}
\end{table}

All PSG recordings include an EMG of the submentalis muscle (chin) and are annotated by sleep experts that detail the sleep stage for every 30 second epoch. Datasets that were annotated using the Rechtschaffen and Kales rules \cite{Rechtschaffen1968} were converted to AASM sleep stages (S3 and S4 were combined and interpreted as N3), which include wake, REM, N1, N2, and N3 \cite{Iber2007}.

Included with these recordings are annotations, that provide movement descriptions along with a timestamp. The descriptors provided are inconsistent and entirely dependent on each sleep technician, they even include spelling errors. All recordings are provided with EMG electrodes placed on the left and right tibias (TIBL and TIBR, respectively). Consequently, this study focused on descriptors that detail leg movements, where examples of text are detailed in bold in Table \ref{table:descriptors}. 

\begin{table}[]
\caption{A list of descriptors detailing movement in the polysomnography recordings. Text in bold are identified as leg limb movement based on text.}
\begin{tabular}{ll}
\hline
\multicolumn{2}{c}{Descriptors}                                       \\ \hline
1. Arousal                        &26.  mouthing and arm movements         \\
2. EVENT 6                      &27. move arms                          \\
3. EVENT5                       &28. move both arms                     \\
4. Event 11                     &29. \textbf{move foot}                          \\
5. Event 15                     &30. move hands                         \\
6. Event 16                     &31. move head                          \\
7. Event 17                     &32. \textbf{move head and legs}                 \\
8. Event 19                     &33. move head and right arm            \\
9. Event 20                     &34. move head from side to side        \\
10. Event 21                     &35. move left                          \\
11. Event 22                     &36. move left arm                      \\
12. Event 4                      &37. \textbf{move legs}                          \\
13. Event 7                      &38. move limb                          \\
14. Event 9                      &39. move right arm                     \\
15. Limb Movement                &40. moveing arm                        \\
16. arm                          &41. moving hands                       \\
17. arm movements                &42. moving head                        \\
18. event 23                     &43. shft positon                       \\
19. fine movements of head       &44. \textbf{shift legs}                         \\
20. good range o jerks           &45. shift position                     \\
21. good range of jerks          &46. shifting limbs                     \\
22. hand fiddling                &47. shifting position                  \\
23. head moves from side to side &48. \textbf{small twitches leading to leg jerk} \\
24. head twitch                  &49. \textbf{straighten legs}                    \\
25. \textbf{lwg twitch}                   &50. twichy hands                      \\ \hline
\end{tabular}
\label{table:descriptors}
\end{table}

\section{Data Processing and Model Training}
\subsection{Signal Preprocessing}
All EMG signals from participants were re-sampled at $256$Hz and filtered between $10$ and $100$Hz (as this is the expected EMG frequency spectrum \cite{Boxtel1984}), using an $8$\textsuperscript{th}-order bandpass filter. Finally a $10$\textsuperscript{th}-order $50$Hz notch filter was also used to suppress noise from mains supply. 
\subsection{Movement Window Size}
% Data driven approach to understanding movement window
While this dataset provided manual annotations of limb movements with a given time-stamp, there is no detail on the duration of the movement. The AASM ascribes limb movement duration varies between $0.50$ and $10$ seconds \cite{Iber2007}. Motivated by a data-driven approach, this study sought to identify all unique annotations during REM sleep and to manually verify annotations that clearly describe leg limb movements. A distribution of absolute amplitude values $10$ seconds before and after the annotation indicated that the majority of activity occurred on average two seconds before and $10$ seconds after the annotated time-stamp. As a result features extracted for the purposes of this study in order to detect leg-movement focused on $10$ second windows. 

\subsection{Feature Extraction}
From each $10$ second window numerous features were calculated in order to train models to understand leg-movement and the absence of leg-movement. These include commonly used features that describe visual characteristics, such as maximum amplitude (Amax), mean amplitude (Amean), standard deviation (Astd), variance, and the 75\textsuperscript{th} percentiles. Another popular feature used was the average power between $10-50$Hz, which was calculated by integrating (rectangular method) the power spectral density function. EMG energy, as described by Liang \textit{et al.} 2012, was also extracted and measures the mean absolute amplitude over each mini-epoch in order to quantify body movement \cite{Liang2012}. The entropy of each mini-epoch was also calculated, which measures the variability of the distribution of the amplitude values \cite{Moddemeijer1989}. The EMG relative spectral power (RSP) was also calculated for frequencies between $10$-$12$Hz (RSP alpha), $12$-$30$Hz (RSP beta), and $30$-$40$Hz (RSP gamma). Additional features included commonly used metrics for evaluating EMG signals to detect the absence of atonia. 

These features included the spectral edge frequency, defined as the frequency below which $95\%$ of the signal power is contained \cite{Imtiaz2014}. The atonia index was also calculated for each mini-epoch, which has been associated with RBD identification since 2008 and was further improved in 2010 \cite{Ferri2008,Ferri2010}. The quantified motor activity (QMA) technique was also used to extract the QMA amplitude, QMA baseline, QMA duration, and the QMA percentage from each mini-epoch. The fractal exponent was also extracted, which measures signal complexity by fitting a linear line to a double logarithmic graph of spectral power density versus frequency \cite{Krakovska2011}. Our previous work has demonstrated the utility of the fractal exponent in RBD detection \cite{Cooray2019}. Finally the manually annotated sleep stage was also added as a feature to focus models to identify movement during REM stages of sleep.  

\subsection{Feature Selection} \label{feat_sel}
It was prudent to utilise feature selection algorithms to identify the most parsimonious set of features to train an effective leg-movement detection classification model. This study employed the minimum redundancy - maximum relevance (mRMR) feature selection algorithm, through the calculation of mutual information \cite{Peng2005}.

\subsection{Classification}
%DPGMM - Zheng
This study chose a Dirichlet Process (DP) mixture model to classify leg-movements based on EMG features. This section details the DP framework and how extracted features are used to form two distributions, describing leg-movement and no leg-movement. These distributions ares modelled by two Gaussian mixture models (GMM), with a DP as a prior. This work was inspired by  the success of this classification in the sleep apnea detection using oxygen saturation data as detailed by Li \textit{et al.} 2019 \cite{Li2020}.

\subsubsection{Movement Detection from a Dirichlet Process Mixture Model}
A selected number of features, as described in Section \ref{feat_sel}, are extracted from segments that have leg-movement and no leg-movement. The classification of these segments can be analysed by comparing the probability of each segment being generated from models of 'leg-movement' and 'no leg-movement'.   

The distributions of features from 'leg-movement' and 'no leg-movement' segments can be modelled by two Gaussian mixture models (GMMs), as a GMM can approximate any distribution accurately by setting an appropriate number of components and adjusting parameters. For this study the two GMM models are the same but are trained using different segments, those from 'leg-movements' and 'no leg-movements'.

Training data, $\boldsymbol{X}$, consisted of features from 'leg-movements', $\boldsymbol{X^1}=\{x_i^1\}_{i=1}^{N_1}$, and 'no leg-movements', $\boldsymbol{X^0}=\{x_i^0\}_{i=1}^{N_0}$. The probability of testing data, $x'$, being generated from either model can be calculated using \eqref{eq18}. Finally a mini-epoch can be classified as 'leg-movement' by:
\begin{align}
\log \frac{p(\boldsymbol{x}'\mid \boldsymbol{X}^1)}{p(\boldsymbol{x}'\mid \boldsymbol{X}^0)}\geq c.\label{eq19}
\end{align}
where $c$ is the threshold for classification, influencing the balance of sensitivity and specificity. This was shown to be effective in a study on apnea detection \cite{Li2020}.
While the idea of independently control for each limb seems trivial, the literature on independent limb movement is not definitive. Studies in human locomotion have demonstrated various degrees of dependence and relative independence \cite{Swinnen1988}. This is further compounded by the question of independent limb movement during sleep, but for the purposes of this study we have assumed that they are independent. Therefore, features derived from the left and right limb electromyogram sensors can be considered independent sources and the log-likelihood can be expressed as follows:

\begin{align}
p(\boldsymbol{x}'\mid\boldsymbol{X}) &= p(\boldsymbol{l}',\boldsymbol{r}'\mid\boldsymbol{L},\boldsymbol{R}) \\ &=p(\boldsymbol{l}'\mid\boldsymbol{L})\cdot p(\boldsymbol{r}'\mid\boldsymbol{R}) \\
\log \frac{p(\boldsymbol{l}'\mid \boldsymbol{L}^1)}{p(\boldsymbol{l}'\mid \boldsymbol{L}^0)}&+\log\frac{p(\boldsymbol{r}'\mid \boldsymbol{R}^1)}{p(\boldsymbol{r}'\mid \boldsymbol{R}^0)}\geq c.\label{eq20}
\end{align}
where training data $\boldsymbol{L}$ and $\boldsymbol{R}$, consisted of features from left and right limb sensors, respectively. While testing data $\boldsymbol{l'}$ and $\boldsymbol{r'}$ are from left and right sensors, respectively. Using cross-fold validation the $c$ threshold was optimised based on the F1-score. 

\section{Results \& Discussion}
\begin{figure*}[htbp]
\centerline{\includegraphics[width=1.0\textwidth]{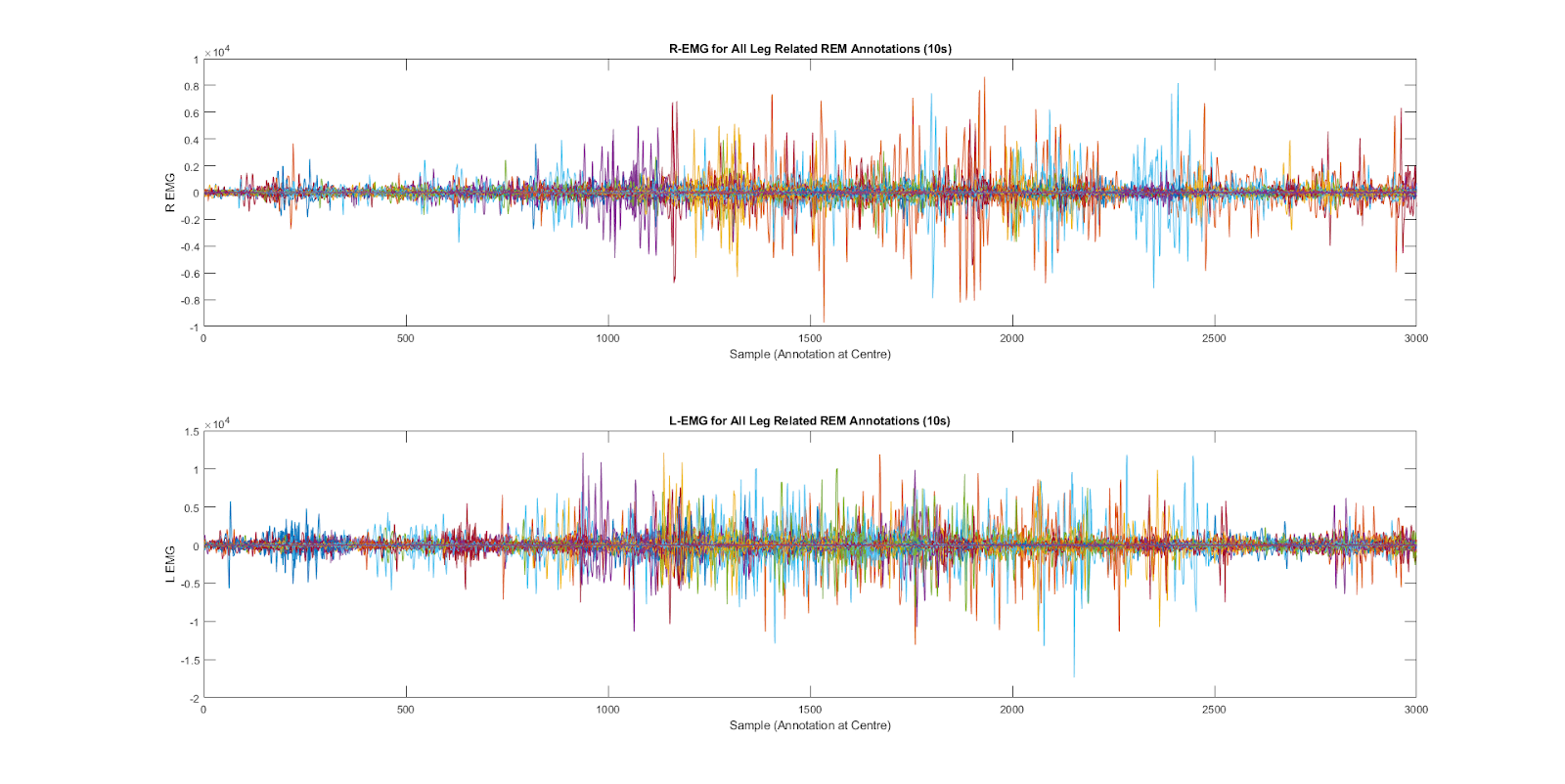}}
\caption{This figure illustrates the signal amplitude from the left and right limb electromyogram in the period ten seconds before and after a leg-movement annotation (provided by sleep clinicians).}
\label{fig1}
\end{figure*}

Using the LEMG and REMG signals available in the PSG recordings described in Section \ref{Data}, an overlay of all limb-movement annotations are detailed in Figure \ref{fig1} (ten seconds before and after an annotation). From this figure we can observe that most amplitude activity occurs two seconds before and eight seconds after an annotation of leg-movement. This attribute informed the decision to extract features from 10 second mini-epochs. These features were used to train and test the DPGMM to detect mini-epochs with leg-movements through a 10-fold cross-validation scheme.

The results of 'leg-movement' detection using the DPGMM are detailed in Table \ref{table:results} along with classification from a random forest model. The DPGMM provides superior precision and F1-score, but achieves a smaller sensitivity compared to the random forest model. The relatively low sensitivity might be due to the wide distribution of features for mini-epoch with and without leg-movement. As a result the trained model becomes sensitive to mini-epochs with strong activity indicative of leg-movement and was unsuccessful at classifying mini-epochs with small segments of movement activity. Nonetheless, the DPGMM was able to achieve a mean precision of $0.25$ and a mean specificity of $0.95$. While this performance might not be able to identify all leg-movements, its precision and specificity might mean this technique is effective at detecting movement for the purposes of RBD identification and diagnosis. 
As instances of leg-movement have a wide spectrum with respect to EMG amplitude activity (for each episode and for every participant), this becomes the underlying cause of misclassification. The DPGMM outperforms the random forest model because it can take into account the features describing different levels of activity when estimating distributions and their Gaussian components based on training data. Once more the features that optimised the DPGMM can be analysed to identify important features in leg-movement detection. While movement during REM sleep constitutes a major criteria for diagnosing RBD, leg movement, specifically, might not be the most frequent \cite{Frauscher2008,Stefani2015}. However, this application of targeting leg movement for RBD participants provides a proof-of-concept that could be applied to other limbs and sleep disorders.  

\begin{table}[htbp]
\caption{Results of leg-movement detection using a random forest (RF) model compared to the Dirichlet process Gaussian mixture (DPGMM) model.}
\begin{center}
\resizebox{\linewidth}{!}{
\begin{tabular}{|l|l|l|l|l|l|}
\hline
              & \textbf{Accuracy} & \textbf{Sensitivity} & \textbf{Specificity} & \textbf{Precision} & \textbf{F1}   \\ \hline
\textbf{RF} & $0.90\pm0.028$     & $0.79\pm,0.12$        & $0.90\pm0.03$        & $0.17\pm0.058$      & $0.27\pm0.082$ \\ \hline
\textbf{DPGMM}         & $0.94\pm0.033$     & $0.48\pm0.19$        & $0.95\pm0.037$        & $0.25\pm0.14$      & $0.30\pm0.15$ \\ \hline
\end{tabular}}
\label{table:results}
\end{center}
\end{table}

\begin{figure}
     \centering
    \begin{subfigure}[b]{0.5\textwidth}
        \includegraphics[width=\textwidth]{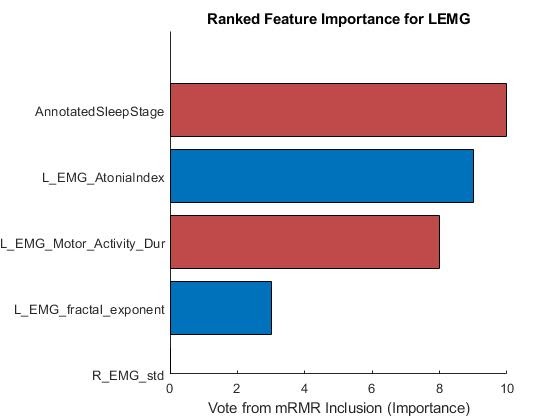}
        \caption{Left limb features.}
        \label{fig2a}
    \end{subfigure}
    \vfill
    \begin{subfigure}[b]{0.5\textwidth}
        \includegraphics[width=\textwidth]{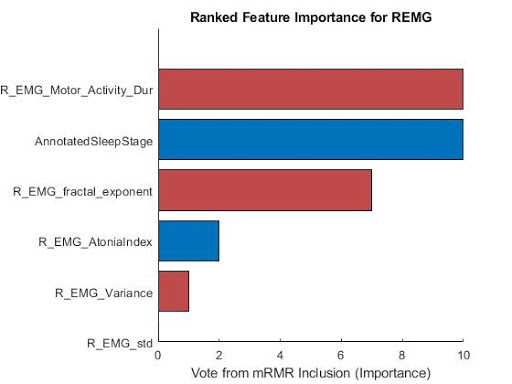}
        \caption{Right limb features.}
        \label{fig2b}
    \end{subfigure}
\caption{The number of instances each feature was used for a Dirichlet Process Gaussian model as determined by a minimum redundancy maximum relevance feature selection algorithm for (a) left limb feature and (b) right limb features.}
\label{fig2}
\end{figure}

During the feature selection process (part of cross-validation), the number of instances when each feature was included ('votes') in the trained model are detailed in Figure \ref{fig2} as a proxy for feature importance. It is clear to see similarities between the importance of left and right limb features, where the annotated sleep stage was the most important for both models of the left and right limbs. This is not surprising that the leg-movement annotations were only identified for REM sleep, resulting in a model focused on the feature of annotated sleep stage. In this study manually annotated sleep staging was already provided, but remains an arduous and time-consuming process, which would hamper any automated process to detect leg-movements and in-turn individual with RBD or PLMD. Additionally, important features also included the atonia index, motor activity (duration), and fractal exponent. These features are prominent because they are able to quantify EMG activity effectively and are more robust to noise. 

\begin{figure*}[htbp]
\centerline{\includegraphics[width=1.0\textwidth]{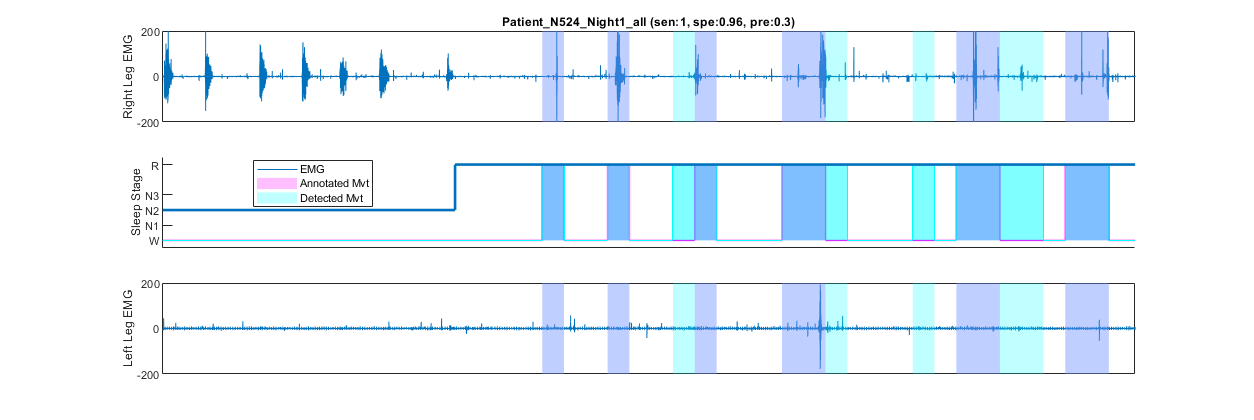}}
\caption{This figure details the left and right leg EMG signal along with the annotated sleep stage with period of annotated and detected movement shaded in magenta and cyan, respectively. Where periods of leg movement annotations have been successfully detected, the period are highlighted in blue (a crossover between cyan and magenta). From this example we can observe the over-sensitive movement detection, where periods of cyan occur at positions corresponding to slight electromyogram (EMG) movement but was not annotated as leg-movement. These periods could coincide with body movement that inevitably are detected by leg EMG sensors.}
\label{fig3}
\end{figure*}

A visual representation of the DPGMM leg-movement detection algorithm is depicted in Figure \ref{fig3}. It is clear to see from this example the left and right leg EMG signals provides information to detection leg-movement. However, from this example we can also observe how slight perturbations in the EMG signal can cause false-positives, reducing the precision of the algorithm.

This study could be further validated by incorporating additional data from healthy control participants and those with other sleep disorders. Furthermore, these leg-movement detection results could provide metrics to identify individuals with specific sleep disorders such as RBD and PLMD. While annotated data for sleep movement is limited and difficult to source, the potential to explore unsupervised methods and the application of transfer learning may prove fruitful. Furthermore the utilisation of GMMs provides the ability to analyse uncertainty assessments, which would provide an interesting future extension of this work. Future work might also look towards including video data \cite{Zhang2008} or utilising non-contact ultrasound Doppler sensors \cite{Fu2021} for the purposes of leg-movement detection or more general movement detection. A further extension of this work could look to incorporate automatic sleep staging to avoid time-consuming and laborious manual sleep staging, providing a much more viable automated diagnostic tool. 

%\begin{figure}[htbp]
%\centerline{\includegraphics[width=0.5\textwidth]{RF_FeatImp.jpg}}
%\caption{This figure.}
%\label{fig4}
%\end{figure}

\section{Conclusion}
The proposed framework described in this study was able to effectively identify leg-movement activity in a dataset of participants diagnosed with RBD by fusing EMG sensors from the left and right limb. To classify leg-movement mini-epochs, four GMMs are trained using features from left and right sensors and from mini-epochs containing 'leg-movement' and 'no leg-movement'. All parameters are derived from the training data by setting the prior of the GMMs as DPs. The most important features as determined by the mRMR feature selection algorithm was the annotated sleep stage, atonia index, motor activity (duration), and the fractal exponent. Future work will look to utilise these models to identify participants with specific sleep disorder, while incorporating additional datasets, and the inclusion of other features from video data.

\section*{Acknowledgment}
The work reported here was sponsored by Research England’s Connecting Capability Fund award CCF18-7157 - promoting the Internet of Things via Collaboration between HEIs and Industry (Pitch-In). This research was also supported by Research Council UK (RCUK) Digital Economy Programme (Oford Centre for Doctoral Training in Healthcare Innovation - grant EP/G036861/1, Sleep and Circadian Neuroscience Institute (SCNi), National Institute for Health Research (NIHR) Oxford Biomedical Research Centre (BRC) and the Engineering and Physical Sciences Research Council (EPSRC – grant EP/N024966/1). We are grateful to EPSRC for funding this work through EP/T013265/1 project NSF-EPSRC: "ShiRAS. Towards Safe and Reliable Autonomy in Sensor Driven" and the support for ShiRAS by the National Science Foundation under Grant NSF ECCS 1903466. The content of this article is solely the responsibility of the author and does not necessarily represent the official views of the RCUK, SCNi, NIHR, EPSRC, or BRC.

\bibliographystyle{IEEEtran2}
\bibliography{RBDBib}

\vspace{12pt}

\end{document}